\documentclass[pra, 12 pt]{revtex4}
\usepackage[english]{babel}
\usepackage{amsmath}
\usepackage{epsfig}

\begin{document}

\title{\bf KRAMERS-KRONIG RELATIONS FOR THE DIELECTRIC FUNCTION AND THE STATIC CONDUCTIVITY OF COULOMB SYSTEMS
}
\author{V.B. Bobrov $^1$, S.A. Trigger $^1$, G.J.F. van Heijst $^2$, P.P.J.M. Schram $^2$}
\address{$^1$ Joint\, Institute\, for\, High\, Temperatures, Russian\, Academy\,
of\, Sciences, 13/19, Izhorskaia Str., Moscow\, 125412, Russia;\\
emails:\, vic5907@mail.ru,\;satron@mail.ru\\
$^2$ Eindhoven  University of Technology, P.O. Box 513, MB 5600
Eindhoven, The Netherlands}

\begin{abstract}
 The mutual influence of singularities of the dielectric permittivitty $\varepsilon(q,\omega)$ in a Coulomb system in two limiting cases $\omega\rightarrow 0$, $q\rightarrow 0$ and $q\rightarrow0$, $\omega\rightarrow0$ is established. It is shown that the dielectric permittivity $\varepsilon(q\rightarrow 0,\omega)$ satisfies the Kramers-Kronig relations, which possesses the a singularity due to a finite value of the static conductivity. This singularity is associated with the long "tails"\, of the time correlation functions.\\

PACS number(s): 05.20.Dd, 05.30.Fk, 52.25.Mq, 72.10.Bg\\

\end{abstract}

\maketitle

\section{Introduction}

The dielectric function $\varepsilon(q,\omega)$ of a Coulomb system is one of the most important characteristics of matter [1-3]. Exact relations for the dielectric function (dielectric permittivity, abbreviated below as DP), which determine general behavior of the DP as a function of the wave vector ${\bf q}$ (in the present paper we consider only the isotropic Coulomb system, therefore the vector ${\bf q}$ is replaced by its modulus $q$) and the frequency $\omega$, have a special significance in the theory of Coulomb systems (CS). At first, it is necessary to mention the Kramers-Kronig relations (KKR).
According to the theory of linear response [2,3], the DP $\varepsilon(q,\omega)$ of a homogeneous and isotropic Coulomb system in volume $V$ at temperature $T$ is determined for $q\neq 0$ by the expression

\begin{eqnarray}
\frac{1}{\varepsilon(q,\omega)}=1+\frac{4\pi}{q^2}\chi^R(q,\omega+i0), \label{B1}
\end{eqnarray}
where $\chi^R(q,z)$ is the retarded "charge-charge" correlation function,
analytical in the upper half-plane of the complex variable $z$ ($Im z>0$):
\begin{eqnarray}
\chi^R(q,z)=-\frac{i}{\hbar V}\int^\infty_0 dt \,\exp(i z t)<[
\rho_q (t),\rho_{-q} (0)]>=\frac{1}{V}<<\rho_q \mid
\rho_{-q}>>_{z}. \label{B2}
\end{eqnarray}
Here $\rho_q(t)$ is the Fourier-component of the charge density in the Heisenberg representation, and the angle brackets $<...>$ denote the averaging in the grand
canonical ensemble, containing charge particles of various species $a$ with charge $z_a e$, mass $m_a$,  chemical potential $\mu_a$ and average density $n_a$. We assume that the quasi-neutrality condition
\begin{eqnarray}
\sum_a e z_a n_a=0 \label{B3}
\end{eqnarray}
is satisfied. The equality \;(\ref{B2})\; has to be treated in the thermodynamic limit\;\; $V\;\rightarrow\;\infty$,  $<N_a> \, \rightarrow \infty$, \;\;\;$n_a\;\;=\;\;<N_a>/V \rightarrow constant$, with $N_a$ the operator of the total number of particles of species $a$. On the basis of Eqs. (\ref{B1})-(\ref{B2}) and the spectral representation for the function $\chi^R(q,\omega+i0)$ it is easy to check the validity of the relations [1-3]:
\begin{eqnarray}
Re \varepsilon(q,\omega)=Re \varepsilon(q,-\omega),\;\; Im
\varepsilon(q,\omega)=-Im \varepsilon(q,-\omega), \label{B4}
\end{eqnarray}
\begin{eqnarray}
Im \varepsilon(q,\omega)>0 \;\;\; \mbox{for}\;\; \omega>0,
\label{B5}
\end{eqnarray}
\begin{eqnarray}
\varepsilon^{-1}(q,0)<1.
\label{B6}
\end{eqnarray}
and also the KKR for the inverse DP $\varepsilon^{-1}(q,\omega)$:
\begin{eqnarray}
Re\, \varepsilon^{-1}(q,\omega)=1+P \int_{-\infty}^\infty \frac{Im
\varepsilon^{-1}(q,\xi)}{\pi}\frac{d \xi}{\xi-\omega}, \label{B7}
\end{eqnarray}
\begin{eqnarray}
Im\, \varepsilon^{-1}(q,\omega)=-P \int_{-\infty}^\infty \frac{[Re
\varepsilon^{-1}(q,\xi)-1]}{\pi}\frac{d \xi}{\xi-\omega}. \label{B8}
\end{eqnarray}
The symbol $P$ in (\ref{B7})-(\ref{B8}) means that we consider the principal value of the integral.
It should be stressed that the KKR can be written in the form (\ref{B7})-(\ref{B8}) on the condition that the functions $\varepsilon^{-1}(q,\omega)$ and $\varepsilon(q,\omega)$ have no singularities on the real axis $\omega$.
In particular, if the conditions
\begin{eqnarray}
\lim_{\omega\rightarrow 0} Im \varepsilon^{-1}(q,\omega)=0, \;\; \lim_{\omega\rightarrow 0} Im \varepsilon(q,\omega)=0\label{B9}
\end{eqnarray}
are satisfied, it is easy to derive from (\ref{B4}),\,(\ref{B5}) and (\ref{B7}) the inequality (\ref{B6}). The KKR for CS at arbitrary thermodynamic parameters are valid only for the inverse DP $\varepsilon^{-1}(q,\omega)$ [4].
The DP $\varepsilon(q,\omega)$ for arbitrary non-zero wave vectors does not obey to the KKR without violation of the causality and stability conditions [4]. If the KKR are valid for the DP $\varepsilon(q,\omega)$ the value of the static DP $\varepsilon(q,0)$ is restricted by the condition [1,2]
\begin{eqnarray}
\varepsilon(q,0)>1. \label{B10}
\end{eqnarray}
At the same time, according to (\ref{B7}), from the KKR for the inverse DP $\varepsilon^{-1}(q,\omega)$  negative values of the static DP $\varepsilon(q,0)$ are also possible [4]:
\begin{eqnarray}
\varepsilon(q,0)<0. \label{B11}
\end{eqnarray}
Moreover, the inequality (\ref{B10}) is the necessary and sufficient condition of validity of the KKR for the DP [4].
The static DP, determined by the limit
\begin{eqnarray}
\varepsilon(q,0)=\lim_{\omega\rightarrow 0} Re \varepsilon(q,\omega),\label{B12}
\end{eqnarray}
characterizes (taking into account condition (\ref{B9})) the equilibrium state of a Coulomb system in a weak electric field. The frequency dependence of the DP characterizes non-equilibrium processes in CS, connected, at first, with the existence of an electric current. These processes are described by the conductivity $\sigma(q,\omega)$, connected with the DP $\varepsilon(q,\omega)$ by the relation [1,2]
\begin{eqnarray}
\varepsilon(q,\omega)=1+\frac{4\pi i}{\omega+i0}\sigma(q,\omega).\label{B13}
\end{eqnarray}
As is well known, the conductivity characterizes the relation between the current density ${\bf J}(q,\omega)$ and the electric field ${\bf E}(q,\omega)$ in a material: ${\bf J}(q,\omega)=\sigma(q,\omega){\bf E}(q,\omega)$.

The most interesting case is the action of a weakly inhomogeneous electric field ($q\rightarrow0$).
It is easy to see on the basis of physical arguments [1], that the static conductivity is determined by the limit
\begin{eqnarray}
\sigma_{st}=\lim_{\omega\rightarrow 0}\lim_{q\rightarrow 0} \sigma(q,\omega)\geq 0. \label{B14}
\end{eqnarray}
Take now into account that for all known substances, for non-zero temperatures, the static conductivity $\sigma^{(0)}$ has a finite non-zero value. Then from  (\ref{B13}) and (\ref{B14}) it follows that

\begin{eqnarray}
\lim_{q\rightarrow 0}\lim_{\omega \rightarrow 0} \varepsilon(q,\omega)\neq
\lim_{\omega\rightarrow 0}\lim_{q\rightarrow 0} \varepsilon(q,\omega). \label{B15}
\end{eqnarray}

\section{Positivity $\varepsilon(q\rightarrow 0,0)$ and KKR for the DP}

Using the methods of quantum field theory and the diagram technique for the temperature Green functions [3,5], the exact relations for the DP and the response function $\chi^R (q,z)$ can be established. In particular, it was shown in [3] that
\begin{eqnarray}
\chi^R (q,z)=\frac{\Pi(q,z)}{\varepsilon(q,z)};\;\;\;\varepsilon(q,z)=1-u(q)\Pi(q.z), \label{B16}
\end{eqnarray}
where the function $\Pi(q,z)$ is the so-called \,"charge-charge"\,
polarization operator - the irreducible (in the $q$-channel on one line of interaction
$u(q)=4\pi/q^2$) part of the  temperature\, "charge-charge" \,Green function. Therefore, the DP $\varepsilon(q,\omega)$ of a homogeneous and isotropic CS can be represented as
\begin{eqnarray}
\varepsilon(q,\omega)=1-u(q)\Pi(q,\omega+i0). \label{B17}
\end{eqnarray}
It should be mentioned that the polarization operator $\Pi(q,\omega+i0)$ can be introduced without use the perturbation theory of the temperature diagram technique. In contrast with the Green function $\chi^R (q,\omega+i0)$, which determines the response of CS on a weak external field, the polarization operator $\Pi(q,\omega+i0)$ characterizes the response on the full field in a medium. It means in general that the polarization operator is not an analytical function in the upper half-plane of the complex variable $z$. Therefore, as follows from  (\ref{B16}) and (\ref{B17}), the KKR for the DP $\varepsilon(q,\omega)$ can be violated [4]. However, the polarization operator $\Pi(q,z)$, in contrast with the Green function $\chi^R (q,z)$, has no singularities for small values of wave vector $q$, connected with the form of the interaction potential $u(q)=4\pi/q^2$ (in general this statement is true for normal CS). Owing to this circumstance the DP $\varepsilon(q,\omega)$ for small wave vectors satisfies the following limiting relations [6-8]:
\begin{eqnarray}
\lim_{q \rightarrow 0} q^2 \varepsilon(q,0)=4 \pi \sum_{a,\,b} z_a z_b e^2 \left(\frac{\partial n_a}{\partial \mu_b}\right)_{\mu_c,\,T}, \label{B18}
\end{eqnarray}
\begin{eqnarray}
\varepsilon(\omega)=\lim_{q \rightarrow 0} \varepsilon(q,\omega)=1-\frac{\omega_p^2}{(\omega+i0)^2}-\frac{\varphi(\omega+i0)}{(\omega+i0)^2},\label{B19}
\end{eqnarray}
\begin{eqnarray}
\varphi(z)=\frac{4\pi}{3V} \ll I^\beta \mid I^\beta \gg_z, \;\;\; Im z>0, \label{B20}
\end{eqnarray}
\begin{eqnarray}
I^\beta = \sum_a z_a e I_a^\beta,  \;\;\; I_a^\beta=\sum_p \frac{\hbar p_\beta}{m_a}a_p^+ a_p, \; \;\;\; \omega_p=\left(\sum_a \frac{4\pi z_a^2 e^2 n_a}{m_a} \right)^{1/2}.\label{B21}
\end{eqnarray}
Here $I^\beta$ is the operator of the total current, $\omega_p$ is
the plasma frequency.
Using the grand canonical ensemble it is easy to see [9] that
\begin{eqnarray}
T \left(\frac{\partial n_a}{\partial \mu_b}\right)_{\mu_c,\,T} =\frac{1}{V} <\delta N_a \delta N_b>, \;\;\; \delta N_a =N_a - <N_a>. \label{B22}
\end{eqnarray}
Substitution of (\ref{B22}) into (\ref{B18}) leads for arbitrary thermodynamic parameters (in a normal CS) to the following result [10]:
\begin{eqnarray}
\lim_{q \rightarrow 0} q^2 \varepsilon(q,0)=\kappa^2 = \frac{4 \pi}{T}\frac{<Z^2>}{V}\geq 0,\;\;\; Z =  \sum_a z_a e N_a. \label{B23}
\end{eqnarray}
It is evident that the value $\kappa$ (\ref{B10}) in the appropriate limiting cases of weakly interacting classical and weakly interacting degenerate CS corresponds to the inverse Debye and the inverse Thomas-Fermi screening length, respectively [3]. Therefore, for small wave vectors $q$ the static DP of CS satisfies the inequality
(\ref{B10}) $\varepsilon(q\rightarrow 0 ,0)>0$. On this basis we concude that the function $\varepsilon(\omega)$ (\ref{B19}) should satisfy the KKR.

Actually, by integrating (\ref{B20}) by parts, one can rewrite (\ref{B19}) in the form
\begin{eqnarray}
\varepsilon(\omega)=1+\frac{4\pi i \sigma(\omega)}{\omega+i0},\;\;\;\sigma(\omega)=\lim_{q \rightarrow 0} \sigma(q,\omega),\label{B24}
\end{eqnarray}
\begin{eqnarray}
\sigma(z)=-\frac{1}{3V} \ll I^\beta \mid P^\beta \gg_z, \;\;\; P^\beta=\int r^\beta \rho ({\bf r}) d{\bf r}. \label{B25}
\end{eqnarray}
Here the operators $\rho ({\bf r})$ and $P^\beta$ are  the charge density operator and the charge dipole moment operators, respectively, with
\begin{eqnarray}
\frac{d P^\beta (t)}{dt}=I^\beta(t),\;\;\; \frac{4\pi i}{3\hbar V} < [I^\beta, P^\beta]>=\omega_p^2. \label{B26}
\end{eqnarray}
Therefore, we show that the conductivity $\sigma(\omega)$ (\ref{B24})-(\ref{B26}) is determined by the Kubo formula [11] (see also [6,7]).

The functions $z(\varepsilon(z)-1)/4\pi$ and $z^2(\varepsilon(z)-1)/4\pi$ are, according to Eqs. (\ref{B19}),\,(\ref{B20}),\,(\ref{B24}) and (\ref{B25}), analytical functions in the upper half-plane of the complex variable $z$ and satisfy the KKR in the traditional form, similar to (\ref{B7})-(\ref{B8}).
The function $\varepsilon(z)$ is also an analytical function in the upper half-plane of the complex variable $z$. However, the notation of the KKR for this function is different from the traditional form. The reason for this difference is
the singularity $4 \pi i \sigma_{st}/\omega$ of the function $\varepsilon(\omega)$, which lies on the real axis at small $\omega$. This implies that the second condition (\ref{B9}) in the long wavelength limit $q\rightarrow 0$ is violated. Therefore [9], the KKR for the function $\varepsilon(\omega)$ takes the form
\begin{eqnarray}
Re \varepsilon(\omega)=1+P \int_{-\infty}^\infty \frac{Im
\varepsilon(\xi)}{\pi}\frac{d \xi}{\xi-\omega}, \label{B27}
\end{eqnarray}
\begin{eqnarray}
Im \varepsilon(\omega)=-P \int_{-\infty}^\infty \frac{[Re
\varepsilon(\xi)-1]}{\pi}\frac{d \xi}{\xi-\omega}+\frac{4\pi\sigma_{st}}{\omega}. \label{B28}
\end{eqnarray}
Therefore, in the static limit $\omega\rightarrow 0$ we cannot put $\omega=0$ in the integrals in
(\ref {B27}),\,(\ref{B28}) and we arrive at restrictions similar to (\ref{B6}),\,(\ref{B10}) and (\ref{B11}). Moreover, the function $Re \varepsilon(\omega)$ may also possess singularities in the limit $\omega\rightarrow 0$.

\section{Asymptotic behavior of time-dependent correlation functions and the static conductivity}

To analyze the mentioned singularities, let us generalize the relations (\ref{B19}),\,(\ref{B24}) and (\ref{B25}), introducing the functions $\varepsilon^{(0)}(\omega)$, $\sigma^{(0)}(\omega)$ and $\varphi^{(0)}(\omega)$, determined on the real axis $\omega$
\begin{eqnarray}
\varepsilon^{(0)}(\omega)=1+ \frac{4 \pi i \sigma^{(0)}(\omega)}{\omega} =1-\frac{\omega_p^2}{\omega^2}-\frac{\varphi^{(0)}(\omega)}{\omega^2},\label{B29}
\end{eqnarray}
\begin{eqnarray}
\sigma^{(0)}(\omega)=\int_0^\infty \exp (i\omega t) \sigma(t) dt,\;\;\; \varphi^{(0)}(\omega)=\int_0^\infty \exp (i\omega t)\varphi(t) dt,\label{B30}
\end{eqnarray}
\begin{eqnarray}
\sigma(t)=\frac{i}{3 \hbar V}<[
I^\beta(t), P^\beta (0)]>,\;\;\; \varphi(t)=-\frac{i}{3 \hbar V}<[
I^\beta(t), I^\beta (0)]>.\label{B31}
\end{eqnarray}
Taking into account Eq.~(\ref{B26}) it is easy to check that $\sigma(t)$ and $\varphi(t)$ are real functions. The presence of singularities in the function $\varepsilon^{(0)}(\omega)$ is a consequence of behavior of the time correlation functions $\sigma(t)$ and $\varphi(t)$ on a large time scale (in the limit $t\rightarrow\infty$).
If the condition
\begin{eqnarray}
\lim_{t\rightarrow\infty} \int_0^t \sigma(t) dt=0\label{B32}
\end{eqnarray}
is fulfilled, the function $\varepsilon^{(0)}(\omega)$ has no singularities for $\omega\rightarrow 0$. However, actually the condition (\ref{B32}) is violated, since the conductivity $\sigma(\omega\rightarrow 0)$ always has a finite value in a normal CS. Moreover, according to (\ref{B29}) and taking into account (\ref{B26}) we arrive at the equality
\begin{eqnarray}
\lim_{t\rightarrow\infty} \int_0^t \sigma(t) dt=\sigma_{st}=\lim_{t\rightarrow\infty}\frac{i}{3 \hbar V}<[
P^\beta(t), P^\beta (0)]>. \label{B33}
\end{eqnarray}

It should be stressed that in the classical limit ($\hbar\rightarrow0$) Eq.~(\ref{B33}) can be represented in a form equivalent to the Einstein relation [12] for the diffusion coefficient of a Brownian particle
\begin{eqnarray}
\sigma_{st}=\lim_{t\rightarrow\infty}\frac{1}{3T Vt}
<P^\beta(0) P^\beta (t)>. \label{B34}
\end{eqnarray}

Therefore, the low-frequency expansion ($\omega\rightarrow 0$) of the functions $ \sigma^{(0)}(\omega)$ and
$\varphi^{(0)}(\omega)$ has an asymptotic character. This kind of behavior is possible only if the functions $\sigma(t)$ and $\varphi(t)$ tend to zero at large values of $t$ ($t\rightarrow\infty$), following some non-exponential law. This means that these functions possess so-called \,"long tails". It should be noted that such long tails of correlation functions have also been observed in other systems [12,13] than Coulomb systems.

\section{Conclusions}

In the present paper it is shown that the DP $\varepsilon(q,0)$ of a homogeneous and isotropic Coulomb system has a singularity $\kappa^2/q^2$ (\ref{B23}) for small values $q$ ($q \rightarrow 0$) and satisfies the inequality  $\varepsilon(q\rightarrow 0,0)>1$ for arbitrary thermodynamic parameters. It is shown that this inequality leads to the KKR relations for the DP $\varepsilon(\omega)$, which take the form (\ref{B27}) and (\ref{B28}). For low frequencies the function $\varepsilon(\omega)=\lim_{q\rightarrow0}\varepsilon(q, \omega)$ has a singularity $4\pi i \sigma_{st}/\omega$, which is taken into account in the KKR.

One can assert that there is one-to-one correspondence between these two singularities if the respective coefficients $\kappa$ and $\sigma_{st}$ are non-zero
\begin{eqnarray}
\kappa\neq 0,\;\; \sigma_{st}\neq 0. \label{B35}
\end{eqnarray}

The value $\kappa^{-1}$ in (\ref{B23}) characterizes the penetration depth of the electric field in a medium. As in the case of the\, "metal-dielectric"\, transition, on the basis of a conductivity analysis (see, e.g., [14]), one can assert that the difference between \,"metals"\, and\, "dielectrics"\, has a relative character, since all known dielectrics have a non-zero static conductivity at $T\neq 0$.

The analogous statement is true with respect to the penetration depth $\kappa^{-1}\simeq \sqrt{TV/<Z^2>}$ \,(see (\ref{B15})) of the electric field in a medium. In \,"metals"\, the penetration depth is very small, while for \,"dielectrics"\, the penetration depth can be of the order of the macroscopic size of the system.

According to our consideration one can introduce the concept of the "true" dielectric when the conditions
\begin{eqnarray}
\kappa\rightarrow 0,\;\;  \sigma_{st}\rightarrow 0 \label{B36}
\end{eqnarray}
are satisfied. According to (\ref{B23}), a homogeneous and isotropic Coulomb system is in a "true" dielectric state if the limiting relation
\begin{eqnarray}
\frac{4 \pi<Z^2>}{T V} \rightarrow 0 \label{B37}
\end{eqnarray}
is fulfilled.
On the basis of the above analysis we establish that the static conductivity of a homogeneous and isotropic CS is determined by the "long" tails of the time-dependent correlation functions (\ref{B33}) and (\ref{B34}). According to Eqs. (\ref{B33}) and (\ref{B36}), if the condition  (\ref{B37}) is fulfilled and the CS is in the state of a "true" dielectric, the following limiting relation
\begin{eqnarray}
\lim_{t\rightarrow\infty}\frac{i}{3 \hbar V}<[
P^\beta(t), P^\beta (0)]>\rightarrow 0  \label{B38}
\end{eqnarray}
has to be satisfied.

\section*{Acknowledgment}

The authors (V.B. and S.T.) are thankful to the Netherlands Organization for
Scientific Research (NWO) for support of this work in the
framework of the grant ¹ 047.017.2006.007.

\end{document}